\documentclass[letterpaper]{article} % DO NOT CHANGE THIS
\usepackage{aaai24}  % DO NOT CHANGE THIS
\usepackage{times}  % DO NOT CHANGE THIS
\usepackage{helvet}  % DO NOT CHANGE THIS
\usepackage{courier}  % DO NOT CHANGE THIS
\usepackage[hyphens]{url}  % DO NOT CHANGE THIS
\usepackage{graphicx} % DO NOT CHANGE THIS
\usepackage{natbib}  % DO NOT CHANGE THIS AND DO NOT ADD ANY OPTIONS TO IT
\usepackage{caption} % DO NOT CHANGE THIS AND DO NOT ADD ANY OPTIONS TO IT
\usepackage{algorithm}
\usepackage{algorithmic}
\usepackage{multirow}
\usepackage{amsmath}
\usepackage{tabularx} 
\usepackage{booktabs}
\usepackage{newfloat}
\usepackage{listings}
\DeclareCaptionStyle{ruled}{labelfont=normalfont,labelsep=colon,strut=off} % DO NOT CHANGE THIS
\floatstyle{ruled}
\newfloat{listing}{tb}{lst}{}
\floatname{listing}{Listing}
\title{Towards Harmonization of SO(3)-Equivariance and Expressiveness: a Hybrid Deep Learning Framework for Electronic-Structure Hamiltonian Prediction}
\author{
	Shi Yin\textsuperscript{\rm 1}\thanks{Corresponding authors: shiyin@iai.ustc.edu.cn, helx@ustc.edu.cn},
	Xinyang Pan\textsuperscript{\rm 2},
	Xudong Zhu\textsuperscript{\rm 1,2},
	Tianyu Gao\textsuperscript{\rm 2},
	Haochong Zhang\textsuperscript{\rm 1},
	Feng Wu\textsuperscript{\rm 1,2}, \textit{Fellow of IEEE} and
	Lixin He\textsuperscript{\rm 1,2}\footnotemark[1], \textit{Fellow of IOP}
}

\affiliations {
	\textsuperscript{\rm 1}Institute of Artificial Intelligence, Hefei Comprehensive National Science Center \\
	\textsuperscript{\rm 2}University of Science and Technology of China \\
}

\usepackage{bibentry}
\begin{document}

\maketitle

\begin{abstract}
Deep learning for predicting the electronic-structure Hamiltonian of quantum systems necessitates satisfying the covariance laws, among which achieving SO(3)-equivariance without sacrificing the non-linear expressive capability of networks remains unsolved. To navigate the harmonization between equivariance and expressiveness, we propose a deep learning method, i.e., HarmoSE, synergizing two distinct categories of neural mechanisms as a two-stage encoding and regression framework. The first stage corresponds to group theory-based neural mechanisms with inherent SO(3)-equivariant properties prior to the parameter learning process, while the second stage is characterized by a non-linear 3D graph Transformer network we propose, featuring high capability on non-linear expressiveness. The novel combination lies in the point that, the first stage predicts baseline Hamiltonians with abundant SO(3)-equivariant features extracted, assisting the second stage in empirical learning of equivariance; and in turn, the second stage refines the first stage's output as a fine-grained prediction of Hamiltonians using powerful non-linear neural mappings, compensating for the intrinsic weakness on non-linear expressiveness capability of mechanisms in the first stage. Our method enables precise, generalizable predictions while capturing SO(3)-equivariance under rotational transformations, and achieves state-of-the-art performance in Hamiltonian prediction on six benchmark databases. 
\end{abstract}

	\section{Introduction}

Deep learning methods  \cite{schutt2019unifying,unke2021se,gu2022neural,li2022deep,zhong2023transferable,gong2023general} have emerged as a promising trend for predicting the electronic-structure Hamiltonian, an essential physical quantity in understanding a wide range of properties, including electronic-structures, magnetic properties, optics, transport, and numerous other properties. These methods have offered a way to bypass
the computationally exhaustive self-consistent steps of the traditional Density Functional Theory (DFT) method  \cite{hohenberg1964inhomogeneous,kohn1965self}, thereby
providing a viable pathway for the efficient simulation and design of
large-scale atomic systems, laying the foundation for many down-stream applications \cite{zhang2023artificial} in the information and energy areas.

Despite these progresses, the Hamiltonian prediction task continues to present substantial challenges for deep learning techniques. High numerical accuracy is required to derive reasonable physical quantities, and furthermore, the fidelity of Hamiltonian predictions should not be confined to a specific coordinate system; rather, the results must demonstrate robust covariance and generalizability across various choices of reference frames. However, achieving 3D rotational equivariance, i.e., equivariance to the SO(3) group, is a tough target for a deep learning Hamiltonian prediction method. This difficulty arises because the Hamiltonian of each pair of atoms is usually high-dimensional, and its variation space under rotational disturbance is large. Consequently, it is difficult to cover the vast variability space they inhabit with rotations merely depending on parameter learning from discrete training samples. To address this, several works \cite{e3nn,gong2023general} applied group theory-guided feature descriptors and tensor operators assuring inherent SO(3)-equivariance prior to the data-driven parameter learning process. Yet, to guarantee such SO(3)-equivariance independent to specific network parameters, these methods highly restricted the use of non-linear activation layers for SO(3)-equivariant features, leading to bottlenecks in expressiveness for complex non-linear mappings, limiting the accuracy achievable in predicting Hamiltonians. This dilemma, is also broadly prevalent in other 3D machine learning tasks where SO(3)-equivariance is highlighted, as analyzed by \citet{DBLP:conf/nips/ZitnickDKLSSUW22}.

To harmonize SO(3)-equivariance and expressiveness for the prediction of electronic-structure Hamiltonians, this paper proposes a two-stage encoding and regression framework, i.e., HarmoSE,  which combines mechanisms boasting parameter-independent prior SO(3)-equivariance with mechanisms featuring flexibility in non-linear expressiveness, overcoming the respective challenges of each categories of mechanisms, i.e., the limited non-linear expressive capability for the former as well as the difficulty of learning SO(3)-equivariance from data for the latter, through effective complementary strategies. Specifically, the first stage corresponds to the neural mechanisms constructed based on group theory with prior equivariant properties of 3D atomic systems, predicting an approximate value of the Hamiltonian, with abundant SO(3)-equivariant features provided.  In the second stage, a highly expressive graph Transformer network we design, with no restrictions on non-linear activations, takes over. This network dynamically learns the 3D structural patterns of the atomic systems, compensates for the expressiveness shortcomings of the first-stage network arising from limited non-linear mappings, and refines the Hamiltonian values predicted in the first stage to enhance accuracy. Although this stage might not possess a parameter-independent prior SO(3)-equivariance due to its non-linearity, it is capable of capturing SO(3)-equivariance through  learning effective network parameters with the help of three pivotal mechanisms. First, instead of directly regressing the entire  Hamiltonians, the second stage aims to refine the Hamiltonian predictions from the first stage with corrective adjustments in a cascaded manner. The scopes of adjustments are smaller, lowing down the difficulties on non-linear learning of SO(3)-equivariance. Second, the second-stage network incorporates covariant features, including SO(3)-equivariant features extracted by the first-stage network and SO(3)-invariant features engineered by geometric knowledge, with its inputs to assist in the implicit learning of SO(3)-equivariance. Third, as the core of Transformer, the attention mechanism has potentials to adapt to geometric condition variations such as coordinate transformations, through its dynamic weighting strategy. Collectively, the combination of the two categories of neural mechanisms in the two stages allows the framework to overcome the challenges of each individual mechanism and make precise, generalizable, and SO(3)-equivariant predictions, being much more effective than simply increasing the parameter count for the network from one stage and fine-tuning it alone.  Our method achieves state-of-the-art (SOTA) performance on Hamiltonian regression in six benchmark databases, demonstrating the superiority  of our method. Particularly, our SOTA performance in the twisted samples, which exhibit both  SO(3)-equivariance effects and variations in van der Waals (vdW) interactions due to the inter-layer rotations, comprehensively confirms the robust capability of our model in capturing the intrinsic SO(3)-equivariance of Hamiltonians as well as its strong non-linear expressive power to generalize to complex and dynamic 3D geometric structures of atomic systems.

\section{Related Work}
\label{related}
In this part, we firstly overview deep learning studies on capturing rotational equivariance. After that, we segue into related works on deep Hamiltonian prediction, in which 3D rotational equivariance is pursued.

As representative researches on equivariance to discrete rotational group, Dieleman \textit{et al.} \cite{DBLP:conf/icml/DielemanFK16} introduced cyclic symmetry operations into CNNs to achieve rotational equivariance; Ravanbakhsh \textit{et al.} \cite{DBLP:conf/icml/RavanbakhshSP17}  explored parameter-sharing techniques for equivariance to discrete rotations; Kondor \textit{et al.}  \cite{DBLP:conf/iclr/KondorSPAT18} developed equivariant representations via compositional methods and tensor theory; Zitnick \textit{et al.}  \cite{DBLP:conf/nips/ZitnickDKLSSUW22}, Passaro \textit{et al.}   \cite{DBLP:conf/icml/PassaroZ23} and Liao \textit{et al.} \cite{equiformerv2} introduced spherical harmonic bases for atomic modeling, focusing on rotational equivariance but limited to discrete sub-groups of SO(3)  due to their sampling strategy. These approaches were very effective on discrete symmetries but sub-optimal on handling continuous 3D rotations. Focusing on equivariance to continuous rotational group, Jaderberg \textit{et al.} \cite{DBLP:conf/nips/JaderbergSZK15}  and Cohen \textit{et al.}  \cite{DBLP:conf/iclr/CohenW17}  achieved considerable success in 2D image recognition tasks by modeling equivariance to in-plane rotations. However, their applications were limited within the scope of 2D tasks and did not fit for the more complex demands of equivariance to 3D continuous rotational group, i.e. SO(3), required in the Hamiltonian prediction task. 

In the field of researches on equivariance to SO(3),  approaches like DeepH \cite{li2022deep} explored equivariance via a local coordinate strategy, which made inference within the fixed local coordinate systems built with neighboring atoms, then transferred the output according to equivariance rules to the corresponding global coordinates. However, due to a lack of in-depth exploration of SO(3)-equivariance at the neural mechanism level, this method faced challenges when the local coordinate system underwent rotational disturbances from non-rigid deformation, e.g. the inter-layer twist of bilayer structures. In contrast, methods like TFN \cite{thomas2018tensor}, SE(3)-Transformer \cite{se3former}, E3NN \cite{e3nn}, Equiformer \cite{equiformer}, and DeephE3 \cite{gong2023general}, considered SO(3)-equivariance from the perspective of intrinsic mechanisms of neural networks, developed group theory-informed equivariant operations, such as linear scaling, element-wise sum, direct sum, direct product, and the Clebsch-Gordan decomposition, effectively applying in multiple atomic modeling tasks, where DeepHE3 \cite{gong2023general} stood as a SOTA method across diverse atomic systems on Hamiltonian prediction. However, a common challenge across these methods lies in the fact that, to achieve inherent equivariance prior to the parameter learning process, they forbade the use of complex non-linear mappings like $Sigmoid$, $SiLU$, and $Softmax$ for SO(3)-equivariant features, significantly limiting the network's expressive potential and causing a bottleneck in generalization performance. Although these approaches tried to  enhance expressiveness via a gated activation function, where SO(3)-invariant features undergone through non-linear activation layers were used as gating coefficients that were multiplied with SO(3)-equivariant features, this mechanism, viewed from the perspective of equivariant features, amounted to a linear operation and did not fundamentally improve their expressive capability. For these methods, this equivariance-expressiveness dilemma remains an unsolved problem.

\section{Preliminary}
\label{Preliminary}

In the study of symmetry on mathematical structures, An operation \(A\) is equivariant with respect to \(B\) if applying \(B\) before or after \(A\) has the same effect, expressed as: $A(B(x)) = B(A(x))$. The key equivariance properties of Hamiltonians are the 3D rotational equivariance with respect to reference frame. Specifically, when the reference frame rotates by a rotation matrix denoted as $\mathbf{R}$, the edge of an atom pairs $(i,j)$ transforms from $\mathbf{r}_{ij}$ to $\mathbf{R}\cdot \mathbf{r}_{ij}$, and the Hamiltonians in the direct sum state transforms equivariantly from  $\mathbf{h}_{ij}$ to $D(\mathbf{R}) \cdot \mathbf{h}_{ij}$, where $D(\mathbf{R})$ is the Wigner-D matrix \footnote{Here we present SO(3)-equivariance under the direct sum state due to its simple vector form. For the equivalent formulation under the matrix-formed direct product state, please refer to \citet{gong2023general}}. The requirements for the fitting and generalization capability of a neural network $f_{nn}(\cdot)$ for Hamiltonian prediction can be formally expressed as: $f_{nn}(\{\mathbf{r}_{ij}| i \in Nodes, ij \in Edges \}) \cong \{ \mathbf{h}_{ij} \}$; moreover, the requirement on SO(3)-equivariance   can be represented as: $f_{nn}(\{ \mathbf{R}\cdot \mathbf{r}_{ij} | i \in Nodes, ij \in Edges \}) \cong  \{ D(\mathbf{R}) \cdot \mathbf{h}_{ij}\} $. It is crucial that $f_{nn}(\cdot)$ intrinsically captures SO(3)-equivariance to effectively generalize under rotational reference frames, and meanwhile, $f_{nn}(\cdot)$ must also possess sufficient expressive power to generalize across different types and structures of atomic systems and make accurate predictions.

\begin{figure}[t]
	\centering
	\includegraphics[scale=0.4]{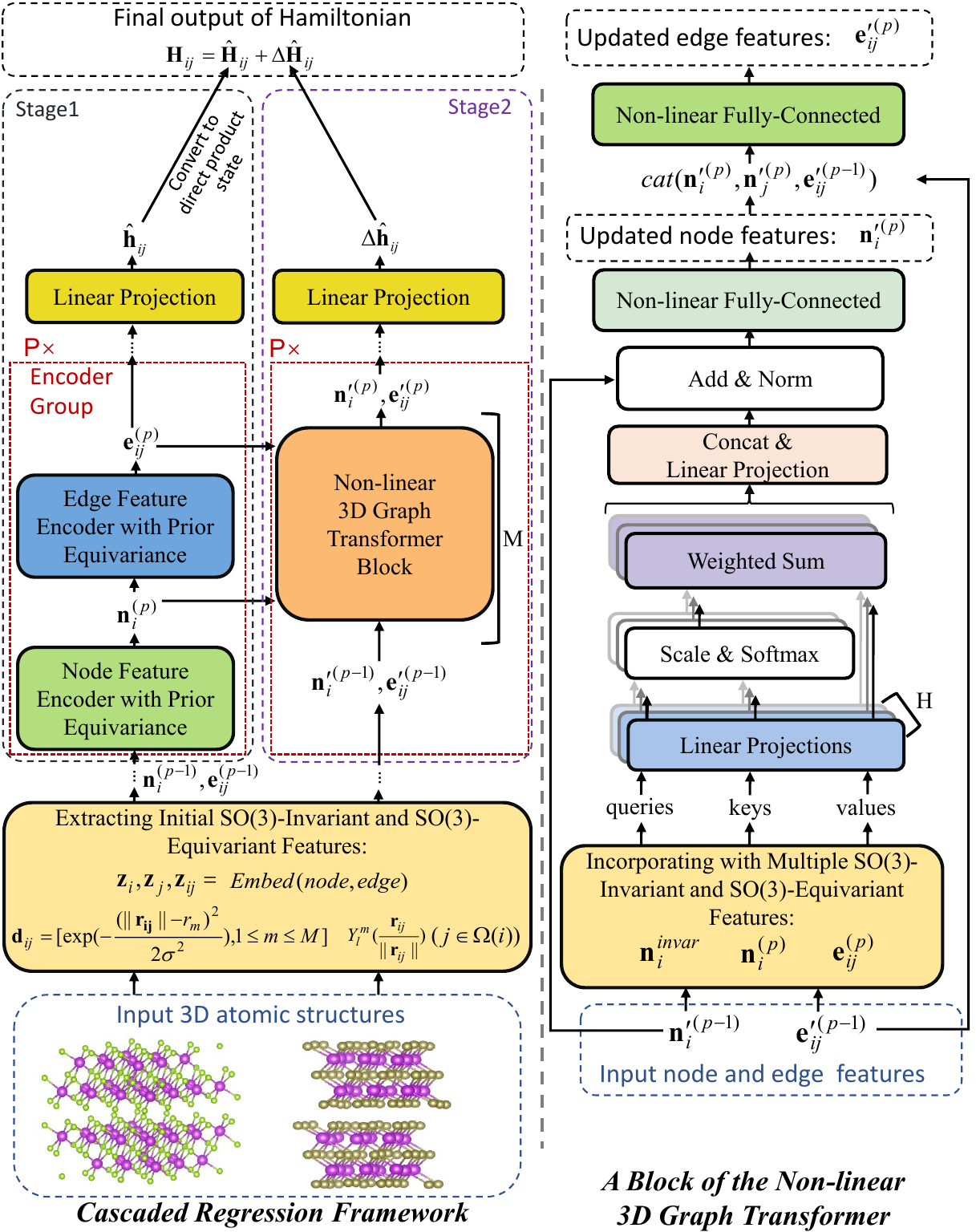}
	\caption{Left part: overview of the two-stage encoding and regression framework for Hamiltonian prediction. Right part: The internal architecture of the non-linear 3D graph Transformer network.}
	\label{framework}
\end{figure}

\section{Method}

As shown in Fig. \ref{framework}, to harmonize SO(3)-equivariance and expressiveness for Hamiltonian prediction, we propose a hybrid framework, i.e., i.e., HarmoSE, with two encoding and regression stages, from which the first-stage network, a group theory-informed network possessing SO(3)-equivariance prior to the learning process, provides essential foundations to the second stage in mastering SO(3)-equivariance, whereas the second-stage network, with highly expressive non-linear mappings, enriches the expressiveness capabilities of the whole framework. The combination of these two stages not only enhances expressiveness but also ensures robust equivariance to rotations of reference frames, bringing accurate predictions for electronic-structure Hamiltonians despite rotational transformations.

\subsection{Initial Features}
\label{ini}
In our framework, the initial feature for the $i (1 \leq i \leq N)$ th node is its node embedding, denoted as $\mathbf{z}_i$, a coordinate-independent SO(3)-invariant semantic embedding that marks its element type. Given the locality of the Hamiltonian \cite{li2022deep}, each atom $j$ in the local set $\Omega(i)$ within the cutoff radius of an atom $i$ form an edge with $i$. From each edge, a Hamiltonian is defined. The initial features for edge $(i, j)$ include both SO(3)-invariant encodings and SO(3)-equivariant encodings. The former includes edge embeddings $\mathbf{z}_{ij}$ marking the types of interacting atom pairs, as well as the distance features $\mathbf{d}_{ij}$ in the form of Gaussian functions \cite{li2022deep}; the latter is spherical harmonics, denoted as $Y_{l}^{m}(\frac{\mathbf{r}_{ij}}{||\mathbf{r}_{ij}||})$ \cite{schrödinger1926quantisierung}, where $\frac{\mathbf{r}_{ij}}{||\mathbf{r}_{ij}||}$ describes the relative orientation between two atoms. 

\subsection{The First Regression Stage}
\label{first}
In our framework, the primary role of the first encoding and regression stage is to extract node and edge representations with intrinsic SO(3)-equivariance independent to network parameters,  and regress baseline Hamiltonian predictions, denoted as $\hat{\mathbf{h}}_{ij}$ in the direct sum state and $\hat{\mathbf{H}}_{ij}$ in the direct product state for each pair $(i,j)$, establishing a firm foundation on SO(3)-equivariance in both feature level and the regression target level. The encoding process of the $p$-th encoder group can be described by the following equation:
\begin{equation}
	\begin{small}
	\label{firststage}
	\begin{aligned}
		\mathbf{n}^{(p)}_i  & = Gate(\mathbf{n}^{(p-1)}_i + \sum_{j \in \Omega(i)}EquiLin(\mathbf{n}^{(p-1)}_i,\mathbf{e}^{(p-1)}_{ij},\mathbf{n}^{(p-1)}_j) \\
		\mathbf{e}^{(p)}_{ij}  & = Gate(EquiLin(\mathbf{n}^{(p)}_i,\mathbf{e}^{(p-1)}_{ij},\mathbf{n}^{(p)}_j)) \\
	\end{aligned}
	\end{small}
\end{equation}
where $\mathbf{n}^{(p)}_i$ and $\mathbf{e}^{(p)}_{ij}$ respectively denote the node and edge features from the $p$ th $(1 \leq p \leq P)$ encoder group, $Gate(\cdot)$ is the gated activation function introduced in the Related Work Section, $EquiLin(\cdot)$ denotes a combination of tensor operators consisting of linear scaling, element-wise sum, direct sum, direct product, as well as the Clebsch-Gordan decomposition, possessing parameter-independent SO(3)-equivariance guaranteed by group theory. These operators, serving as our first-stage backbone, have been comprehensively developed in previous works \cite{thomas2018tensor,e3nn,gong2023general}, achieving maturity in their equivariance capabilities, yet facing inherent limitations in non-linear expressiveness cannot be easily resolved within their mechanisms. Therefore, we focus more on the complementary mechanisms of the hybrid two-stage  encoding and regression framework as well as the design of the second-stage network, aiming at rectifying the intrinsic weaknesses in non-linear expressiveness of the first-stage network while embodying robust SO(3)-equivariant capability. 

%\vspace{-1mm}
\subsection{The Second Regression Stage}
\label{second}
The second encoding and regression stage of our framework is designed to fully exploit non-linear mappings to enhance the expressive capability of the whole framework while capturing  SO(3)-equivariance. For that purpose, as shown in the right part of Fig. \ref{framework}, we propose a 3D graph Transformer which effectively models 3D atomic structures and predicts non-linear correction terms that complement the predictions from the first stage, achieving high-precision Hamiltonian prediction. Yet, one key problem to solve is, non-linear projections might not have parameter-independent guarantee on SO(3)-equivariance, forcing the second stage to capture equivariance through learning effective network parameters from the data. Thus, the difficulty on empirical learning SO(3)-equivariance of Hamiltonians must be addressed. Our second-stage network adeptly resolves this issue, simultaneously capturing SO(3)-equivariance and enhancing the expressive capabilities. This is primarily attributed to three pivotal mechanisms  we design.

First, the second stage works in a cascaded manner for regression, which means that its prediction target is not the entire Hamiltonian but a correction term $\Delta\hat{\mathbf{H}}_{ij}$, relative to the first stage's output, i.e. the initial Hamiltonian estimate $\hat{\mathbf{H}}_{ij}$. The sum of these two stages' outputs forms the final prediction of the Hamiltonian: $\mathbf{H}_{ij} = \hat{\mathbf{H}}_{ij} + \Delta\hat{\mathbf{H}}_{ij}$. Given that the predicted results of $\hat{\mathbf{H}}_{ij}$ are theoretically SO(3)-equivariant and numerically approximate to reasonable, the range of variations for the correction term $\Delta\hat{\mathbf{H}}_{ij}$,  becomes smaller compared to $\hat{\mathbf{H}}_{ij}$. This reduces the complexity of the output space for the second stage and enhances the feasibility on implicitly mastering SO(3)-equivariance through data-driven learning for non-linear modules.

Second, several theoretical-guaranteed covariant features, including both SO(3)-equivariant and SO(3)-invariant features, are integrated into the input features of each Transformer block in the second stage to aid it in capturing equivariance, as shown in Eq. \eqref{megrge_invar}:
\begin{equation}
	\label{megrge_invar}
	\begin{aligned}
		\mathbf{\widetilde{n}}'^{(p)}_i  & = \mathbf{n}'^{(p-1)}_i + \alpha \mathbf{n}^{(p)}_i + \beta  \mathbf{n}^{invar}_i, \\
		\mathbf{\widetilde{e}}'^{(p)}_{ij} & = \mathbf{e}'^{(p-1)}_{ij} +  \lambda \mathbf{e}^{(p)}_{ij}
	\end{aligned}
\end{equation}
where $\alpha$, $\beta$ and $\lambda$ are hyper-parameters, $\mathbf{n}'^{(p-1)}_i$ and $\mathbf{e}'^{(p-1)}_{ij}$ denote the outputs of the Transformer at the $p-1$ ${th}$ encoder group, $\mathbf{\widetilde{n}}'^{(p)}_i$ and $\mathbf{\widetilde{e}}'^{(p)}_{ij}$ respectively serve as the input node and edge features for the subsequent modules of the Transformer at the $p$ ${th}$ encoder group, $\mathbf{n}^{(p)}_i$ and  $\mathbf{e}^{(p)}_{ij}$ are the SO(3)-equivariant node and edge features, respectively, from the corresponding encoder group of the first-stage network. Besides SO(3)-equivariant features, since as demonstrated by literature \cite{wang2018deepmd,zhang2019embedded,external}, SO(3)-invariant features also facilitate the learning of SO(3)-equivariance, we also develop a SO(3)-invariant node feature, i.e.,  $\mathbf{n}^{invar}_i$, aggregated from multiple SO(3)-invariant features, such as node embeddings $\mathbf{z}_i$, edge embeddings $\mathbf{z}_{ij}$, distance features $\mathbf{d}_{ij}$, and triplet angle feature $\theta_{ijk}$ formed by node $i$ as well as two of its local atoms $j$ and $k$, in the way like:
\begin{equation}
	\begin{small}
	\label{megrge_invar1}
	\mathbf{n}^{invar}_i = \sum_{(j,k) \in \Omega(i)} FC(cat(\mathbf{z}_i, \mathbf{z}_j, \mathbf{z}_k, \mathbf{d}_{ij}, \mathbf{d}_{ik}, \mathbf{c}_{ijk}))) 
	\end{small}
\end{equation}
where $cat(\cdot)$ is the concatenation operator, $FC(\cdot)$ denotes fully-connected layers with non-linear activations, $\mathbf{c}_{ijk} = [\cos(\theta_{ijk}), \cos(\theta_{ijk}), ...]$ is a vector extended by duplication, which serves to amplify the angle features for $FC(\cdot)$. To reduce the quadratic complexity, i.e., $O(|\Omega(i)|^2)$  when sampling $(j,k) \in \Omega(i)$, we arrange the set $\Omega(i)$ as an array and only extract adjacent element pairs as tuples $(j, k)$, lowing down the sampling complexity to $O(|\Omega(i)|)$ to efficiently compute $\mathbf{n}^{invar}_i$. In Eq. \eqref{megrge_invar}, $\mathbf{n}^{invar}_i$ is directly merged into node features, and since node features are then merged into edge features in the subsequent modules, it also enhances the learning of edge features. With the help of these covariant features, the second-stage network, even a non-linear one, can also capture the SO(3)-equivariant properties inherent in the Hamiltonian.

Third, we design a multi-head attention mechanism to learn node and edge representations of the 3D atomic systems. The capability to dynamically focus on related geometric features enables robust adaptability to diverse geometric conditions, from structural variants to coordinate transformations. Specifically, the attention mechanism firstly learns dynamic weights, i.e. $\alpha^{(p)}_{ij}$ for the edge $(i,j)$ at the $p$ th encoder group, based on the interactive relationship between the current atom $i$ and its local atoms $j\in \Omega(i)$, as shown in Eq. \eqref{eq:bs1}:
\begin{equation}
	\begin{aligned}
		\label{eq:bs1}
		\textbf{q}^{h(p)}_{ij}  & =   \textbf{W}^h_{q} \cdot cat(\mathbf{\widetilde{n}}'^{(p)}_i,\mathbf{\widetilde{e}}'^{(p)}_{ij}), \\
		\textbf{k}^{h(p)}_{ij}  & =  \textbf{W}^h_{k} \cdot cat(\mathbf{\widetilde{n}}'^{(p)}_j,\mathbf{\widetilde{e}}'^{(p)}_{ij}),\\ 
		\alpha^{(p)}_{ij} &= softmax(\frac{(\textbf{q}^{h(p)}_{ij})^T \cdot  \textbf{k}^{h(p)}_{ij}}{\sqrt{d_h}})
	\end{aligned}
\end{equation}
where $h (1\leq h \leq H)$ is the head index, $d_h$ is the dimension of features, $\textbf{W}^h_{q}$ and $\textbf{W}^h_{k}$ are parameter matrices to calculate queries and keys, i.e., $\textbf{q}^{h(p)}_{ij}$ and $\textbf{k}^{h(p)}_{ij}$, respectively. Here the scale factor $\sqrt{d_h}$ in the denominator is used to prevent $softmax(\cdot)$ from gradient saturation, and the multiple heads aim at enchaining the model capacity. Based on $\alpha^{(p)}_{ij}$, the node features are updated flexibly through the structural information embedded in its local sets, as shown in Eq. \eqref{eq:bs2}:
\begin{equation}
				\begin{small}
	\begin{aligned}
		\label{eq:bs2}
		& \mathbf{v}^{h(p)}_{i}  = \sum_{j \in \Omega(i)}\alpha^{(p)}_{ij} \cdot  ( \textbf{W}^h_{v} \cdot cat(\mathbf{\widetilde{n}}'^{(p)}_j,\mathbf{\widetilde{e}}'^{(p)}_{ij}) ), \\
		& \mathbf{n}'^{(p)}_i  = FC(  LN(\textbf{W}_o \cdot cat(\mathbf{v}^{1(p)}_{i},...,\mathbf{v}^{H(p)}_{i})  + \mathbf{n}'^{(p-1)}_i))
	\end{aligned}
		\end{small}
\end{equation}
where $LN(\cdot)$ is the layer normalization operator, $FC(\cdot)$ denotes fully-connected layers with non-linear activations. Based on $\mathbf{n}'^{(p)}_i$, the edge representations are updated as:
\begin{equation}
	\label{update_e}
	\mathbf{e}'^{(p)}_{ij} = FC(cat(\mathbf{n}'^{(p)}_i, \mathbf{n}'^{(p)}_j, \mathbf{e}'^{(p-1)}_{ij}))) 
\end{equation}
when repeatedly stacking operations in Eq. \eqref{eq:bs2} and \eqref{update_e} in an alternate manner, local patterns can incrementally spread to a larger scale through the neighbors of neighboring atoms. Nevertheless, given the Hamiltonian's locality, there's typically no need for information transfer over very long distances. Finally, the correction term outputted by the second stage is regressed from the edge features $\mathbf{e}'^{(P)}_{ij}$ encoded by the last encoder group, in the way like: 
\begin{equation}
	\Delta\hat{\mathbf{H}}_{ij}  = DStoDP(\Delta\hat{\mathbf{h}}_{ij}) = DStoDP(FC(\mathbf{e}'^{(P)}_{ij})) 
\end{equation}
where $DStoDP(\cdot)$ is the conversion operation from the vector-formed direct sum state to the matrix-formed direct product state, which  is more commonly used in the down-stream computational tasks based on Hamiltonians.

\subsection{Training}
Denote  the ground truth Hamiltonian label for the atom pair $(i, j)$ as $\mathbf{H}^*_{ij}$, in the first stage, parameters of the first-stage network are optimized by minimizing $MSE(\hat{\mathbf{H}}_{ij},\mathbf{H}^*_{ij})$, while in the second stage, parameters of the second-stage network are optimized by minimizing $MSE(\Delta\hat{\mathbf{H}}_{ij},$ $(\mathbf{H}^*_{ij}-\hat{\mathbf{H}}_{ij}))$.  

\section{Experiments}
\subsection{Experimental Conditions}

\begin{figure}
	\centering
	\includegraphics[scale=0.32]{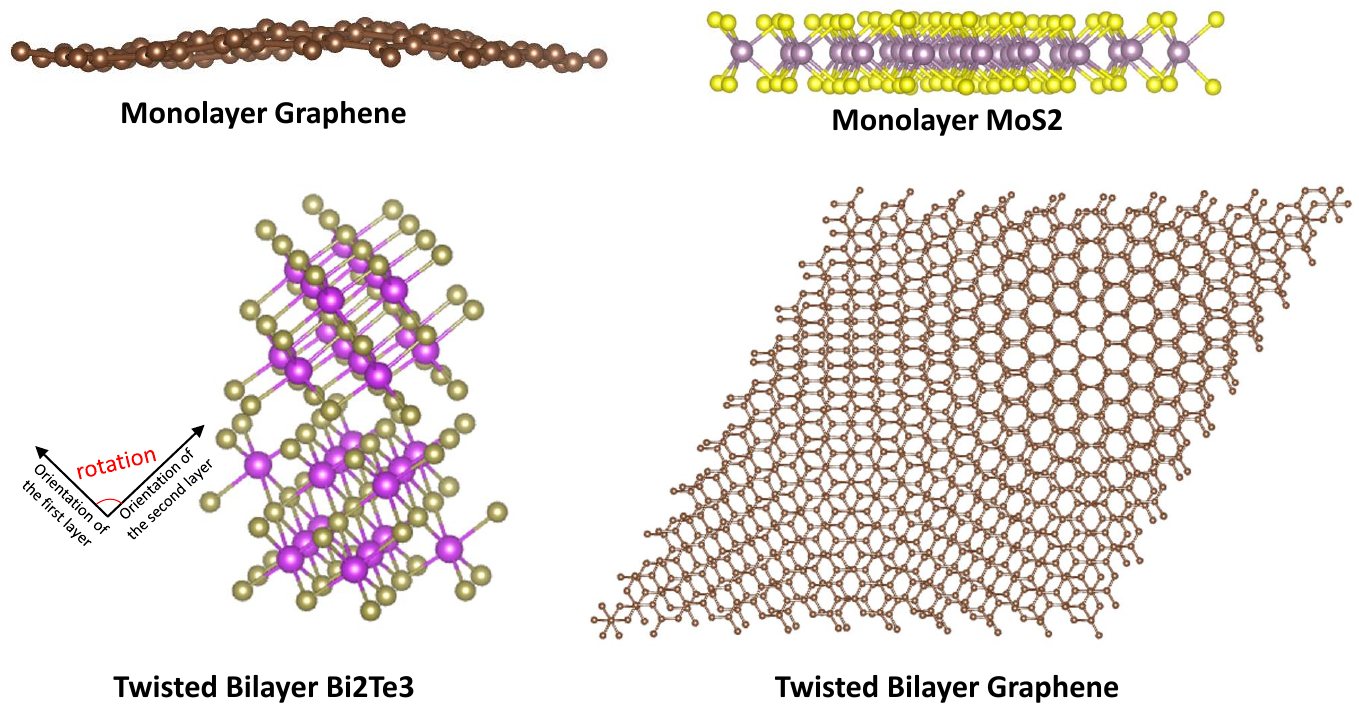}
	\caption{Visualization of challenging testing samples, which exhibit  structural deformations caused by thermal motions and inter-layer twists, calling for strong capabilities on expressiveness and SO(3)-equivariance of a regression model.}
	\label{vis_stru}
\end{figure}

We conduct experiments on six benchmark material databases, including Monolayer Graphene (abbreviated as \textit{MG}), Monolayer MoS2 (\textit{MM}), Bilayer Graphene (\textit{BG}), Bilayer Bismuthene (\textit{BB}), Bilayer Bi2Te3 (\textit{BT}), and Bilayer Bi2Se3 (\textit{BS}), which are released by the DeepH series \cite{li2022deep,gong2023general}.  These databases are diverse and representative, as they cover atomic structures with strong chemical bonds within individual layers and weak vdW interactions between  two layers; and include varied degrees of spin-orbit coupling (SOC), featuring both strong SOC samples like \textit{BT}, \textit{BB} and \textit{BS}, and others with weak SOC. These atomic structures hold significant potential and value in the information science and technology sectors. The training, validation, and testing sets as well as the data pre-processing protocols we use are exactly the same as \cite{gong2023general}. A concise overview of these databases is presented in Table \ref{database-table}. Predicting the Hamiltonian accurately for these atomic structures poses a significant challenge due to the presence of structural deformations caused by thermal motions and inter-layer twists, as shown in Fig. \ref{vis_stru}. Worth noting, the twisted structures have become a research hotspot due to their potentials for new electrical and quantum topological properties \cite{twist,twist2,twist3}. During the twist transformations, the relative rotation between atoms and the coordinate system brings the corresponding SO(3)-equivariant effects; meanwhile, the change in orientations between two layers of atoms causes variations in vdW interactions. These combined effects present a challenge to both of the equivariance and expressiveness capabilities of the regressor. In our experiments, the twisted subsets are even challenging as there are no such samples in the training set. 

\begin{table}
	\caption{The sizes of training, validation, and testing sets including both non-twisted samples (nt) and twisted samples (t),  as well as the dimensions of the Hamiltonian matrices in the direct product state of each atom pair, for each experimental database.}
	\label{database-table}
	\begin{center}
	\begin{small}
		\begin{tabular}{l|c|c|c|ccc}
			\toprule
			& MG & MM & BG & BB & BT & BS \\
			\midrule
			Train (nt) & 270 & 300 & 180 & 231 & 204 & 231\\
			Val (nt) & 90 & 100 & 60& 113 & 38 & 113\\
			Test (nt) & 90 & 100& 60 & 113 & 12 & 113\\
			Test (t) & - & - &9 & 4 & 2 & 2 \\
			$\textnormal{Dim}(\textnormal{\textbf{H}}_{ij})$  &  169 & 361 &169 & \multicolumn{3}{c}{361} \\
			\bottomrule
		\end{tabular}
	\end{small}
	\end{center}
	%\vspace{-3mm}
\end{table}

To ensure the reproducibility, we use a fixed random seed, i.e., $42$,  for all procedures involving randomness, such as parameter initialization, data loader, as well as the rigid rotational augmentation introduced during training. Except for the network modules we need to compare with, i.e., those from DeepHE3 \cite{gong2023general}, the hyper-parameters of our framework are determined based on model selection on the validation sets; for DeepHE3, we adopt the optimal hyper-parameters officially provided, as described below. 

\begin{table*}[t]
	\caption{Comparison of experimental results measured by \(MAE^H_{all}\) as well as \(MAE^H_{cha\_s}\) (meV) on the monolayer structures.}
	\label{results_mono}
	\begin{center}
		\begin{small}
			\setlength{\tabcolsep}{3pt} % Adjusts the space between table columns
			\begin{tabular}{ccccc}
				\toprule
				Method & \multicolumn{2}{c}{MG} & \multicolumn{2}{c}{MM} \\
				\cmidrule(r){2-3} \cmidrule(r){4-5}
				& \(MAE^H_{all}\) & \(MAE^H_{cha\_s}\) & \(MAE^H_{all}\) & \(MAE^H_{cha\_s}\) \\
				\midrule
				$DeepH\!E3$ & 0.251 & 0.357 & 0.406 & 0.574 \\
				$S1_{DeepH\!E3}+S2_{DeepH\!E3}$ & 0.239 & 0.338 & 0.392 & 0.499 \\
				$GFormer$ & 0.816 & 0.897 & 1.025 & 1.250 \\
				$S1_{GFormer}+S2_{GFormer}$ & 0.653 & 0.720 & 0.911 & 0.923 \\
				$S1_{DeepH\!E3}+S2^{-cas}_{GFormer}$ & 0.774 & 0.880 & 0.927 & 0.958 \\
				$S1_{DeepH\!E3}+S2^{-cov}_{GFormer}$ & 0.243 & 0.328 & 0.384 & 0.414 \\
				$S1_{DeepH\!E3}+S2^{-att}_{GFormer}$ & 0.221 & 0.297 & 0.319 & 0.366 \\
				$Ours@(S1_{DeepH\!E3}+S2_{GFormer})$ & \textbf{0.176} & \textbf{0.267} & \textbf{0.233} & \textbf{0.293} \\
				\bottomrule
			\end{tabular}
		\end{small}
	\end{center}
\end{table*}

\begin{table*}[t]
	\caption{Experimental results on the non-twisted subsets (marked with superscripts $nt$) of the bilayer structures.}
	\label{results_bilayer}
	\begin{center}
		\begin{small}
			\setlength{\tabcolsep}{3pt} % Adjusts the space between table columns
			\begin{tabular}{ccccccccc}
				\toprule
				Method & \multicolumn{2}{c}{BG\(^{nt}\)} & \multicolumn{2}{c}{BB\(^{nt}\)} & \multicolumn{2}{c}{BT\(^{nt}\)} & \multicolumn{2}{c}{BS\(^{nt}\)} \\
				\cmidrule(r){2-3} \cmidrule(r){4-5} \cmidrule(r){6-7} \cmidrule(r){8-9}
				& \(MAE^H_{all}\) & \(MAE^H_{cha\_s}\) & \(MAE^H_{all}\) & \(MAE^H_{cha\_s}\) & \(MAE^H_{all}\) & \(MAE^H_{cha\_s}\) & \(MAE^H_{all}\) & \(MAE^H_{cha\_s}\) \\
				\midrule
				$DeepH\!E3$ & 0.389 & 0.453 & 0.274 & 0.304 & 0.447 & 0.480 & 0.397 & 0.424 \\
				$S1_{DeepH\!E3}+S2_{DeepH\!E3}$ & 0.372 & 0.434 & 0.268 & 0.292 & 0.435 & 0.471 & 0.389 & 0.410 \\
				$GFormer$ & 1.295 & 1.483 & 0.886 & 0.949 & 1.018 & 1.230 & 1.352 & 1.483 \\
				$S1_{GFormer}+S2_{GFormer}$ & 0.786 & 0.828 & 0.785 & 0.816 & 0.903 & 0.982 & 0.862 & 0.922 \\
				$S1_{DeepH\!E3}+S2^{-cas}_{GFormer}$ & 0.854 & 0.920 & 0.802 & 0.873 & 0.930 & 1.026 & 0.898 & 0.960 \\
				$S1_{DeepH\!E3}+S2^{-cov}_{GFormer}$ & 0.365 & 0.427 & 0.249 & 0.281 & 0.439 & 0.466 & 0.392 & 0.401 \\
				$S1_{DeepH\!E3}+S2^{-att}_{GFormer}$ & 0.348 & 0.419 & 0.243 & 0.286 & 0.385 & 0.414 & 0.348 & 0.375 \\
				$Ours@(S1_{DeepH\!E3}+S2_{GFormer})$ & \textbf{0.287} & \textbf{0.362} & \textbf{0.172} & \textbf{0.198} & \textbf{0.294} & \textbf{0.321} & \textbf{0.282} & \textbf{0.308} \\
				\bottomrule
			\end{tabular}
		\end{small}
	\end{center}
	%\vspace{-3mm}
\end{table*}
\begin{table*}[t]
	\caption{Experimental results on the twisted subsets (marked with superscripts $t$) of the bilayer structures.}
	\label{results_bilayer2}
	\begin{center}
		\begin{small}
			\setlength{\tabcolsep}{3pt} % Adjusts the space between table columns
			\begin{tabular}{ccccccccc}
				\toprule
				Method & \multicolumn{2}{c}{BG$^{t}$} & \multicolumn{2}{c}{BB$^{t}$} & \multicolumn{2}{c}{BT$^{t}$} & \multicolumn{2}{c}{BS$^{t}$} \\
				\cmidrule(r){2-3} \cmidrule(r){4-5} \cmidrule(r){6-7} \cmidrule(r){8-9}
				& \(MAE^H_{all}\) & \(MAE^H_{cha\_s}\) & \(MAE^H_{all}\) & \(MAE^H_{cha\_s}\) & \(MAE^H_{all}\) & \(MAE^H_{cha\_s}\) & \(MAE^H_{all}\) & \(MAE^H_{cha\_s}\) \\
				\midrule
				$DeepH\!E3$ & 0.264 & 0.429 & 0.468 & 0.602 & 0.831 & 0.850 & 0.370 & 0.390 \\
				$S1_{DeepH\!E3}+S2_{DeepH\!E3}$ & 0.257 & 0.423 & 0.460 & 0.595 & 0.826 & 0.843 & 0.358 & 0.381 \\
				$GFormer$ & 0.982 & 1.153 & 1.784 & 1.921 & 2.682 & 2.827 & 1.785 & 2.037 \\
				$S1_{GFormer}+S2_{GFormer}$ & 0.841 & 0.873 & 1.426 & 1.680 & 2.190 & 2.379 & 1.624 & 1.953 \\
				$S1_{DeepH\!E3}+S2^{-cas}_{GFormer}$ & 0.801 & 0.863 & 1.213 & 1.569 & 1.892 & 1.937 & 1.569 & 1.892 \\
				$S1_{DeepH\!E3}+S2^{-cov}_{GFormer}$ & 0.312 & 0.441 & 0.530 & 0.697 & 0.928 & 0.943 & 0.415 & 0.456 \\
				$S1_{DeepH\!E3}+S2^{-att}_{GFormer}$ & 0.278 & 0.428 & 0.504 & 0.669 & 0.837 & 0.846 & 0.392 & 0.421 \\
				$Ours@(S1_{DeepH\!E3}+S2_{GFormer})$ & \textbf{0.227} & \textbf{0.403} & \textbf{0.438} & \textbf{0.578} & \textbf{0.774} & \textbf{0.794} & \textbf{0.336} & \textbf{0.365} \\
				\midrule
			\end{tabular}
		\end{small}
	\end{center}
	%\vspace{-3mm}
\end{table*}
	\begin{figure*}
	\centering	
	\includegraphics[scale=0.4]{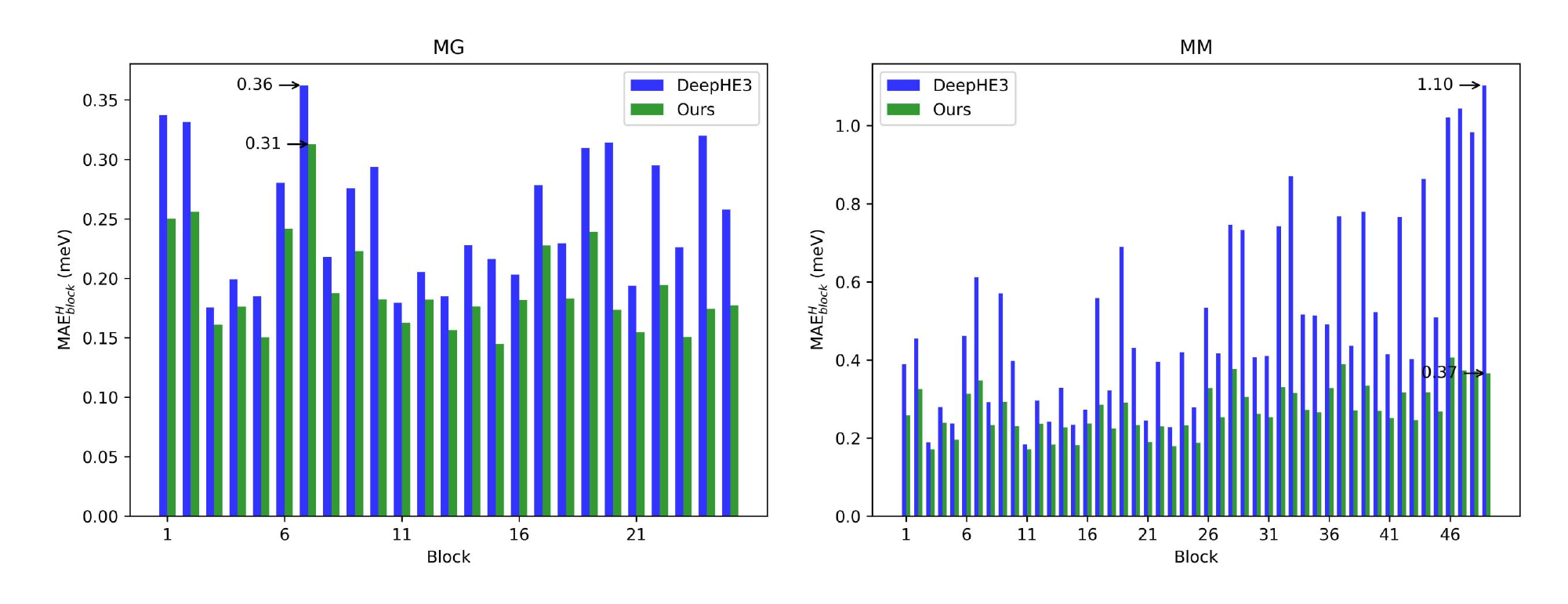}
	\caption{The MAE values (denoted as $MAE^H_{block}$) on various basic blocks of the Hamiltonian matrix in direct product state on the experimental monolayer structures. The MAE values ($MAE^H_{cha\_b}$) on the Hamiltonian block where the baseline model DeepHE3 performs the worst are highlighted for comparison.}
	\label{fig:w}
\end{figure*}
\subsection{Experimental Comparison and Analysis}
\label{eca}
To validate the effectiveness of our approach, we design a set of experimental settings that include not only our complete method but also ablation and comparison settings as follows:

\textbullet\ $DeepH\!E3$ \cite{gong2023general}. This experimental setting evaluates the performance of using only the DeepHE3 architecture to predict Hamiltonians. DeepHE3 will serve as the first-stage network in implementation of our framework in the following experiments. Although our first stage is designed with the flexibility to employ any combination of operators with prior SO(3)-equivariance, for a clear and fair comparison, we here opt to use the  architecture of DeepHE3  which  consists of node and edge encoders with abundant priorly-equivariant operators \cite{e3nn}, under the same hyper-parameters from their open source resources\footnote{https://github.com/Xiaoxun-Gong/DeepH-E3}, ensuring our experimental results can be reliably compared to the established SOTA method. Notably, the results of DeepHE3 reproduced with its latest resources are slightly better than those reported in their original paper.

\textbullet\ $S1_{DeepH\!E3}+S2_{DeepH\!E3}$. This experimental setting arranges two DeepHE3 networks as a two-stage (stage1 and stage2 abbreviated as $S1$ and $S2$, respectively, hereafter) cascaded regression framework for Hamiltonian prediction.

\textbullet\ $GFormer$. This experimental setting evaluates the performance of using only the proposed non-linear graph Transformer (abbreviated as $GFormer$) architecture to predict Hamiltonians. To facilitate the empirical learning of SO(3)-equivariance for non-linear modules, rigid rotational data augmentation on the training samples is introduced. 

\textbullet\ $S1_{GFormer}+S2_{GFormer}$. This experimental setting arranges two of the non-linear graph Transformer networks as a two-stage cascaded regression framework for Hamiltonian prediction.

\textbullet\ $S1_{DeepH\!E3}+S2^{-cas}_{GFormer}$. This experimental setting retains the two-stage encoding framework, only removing the cascaded regression strategy at the output level by directly taking the second stage to  predict the entire Hamiltonian targets. This setup is used to exactly examine the necessity of the cascaded regression strategy, thus only removing the prediction results of the first stage network at the output level, while retaining the first stage's support for the second stage at the feature level.

\textbullet\ $S1_{DeepH\!E3}+S2^{-cov}_{GFormer}$. This experimental setting retains the two-stage regression framework, only removing the mechanism of flowing features with prior covariance into the input layers of Transformer blocks in the second stage. This setup is used to examine the necessity of these covariant features in assisting the non-linear graph Transformer network at learning SO(3)-equivariance. 

\textbullet\  $S1_{DeepH\!E3}+S2^{-att}_{GFormer}$. This experimental setting retains the two-stage encoding and regression framework and their corporation at both feature level as well as output level, only removing the attention mechanism from the Transformer and replacing it by mixing neighboring features by stationary averages similar to DeepHE3. This setup is used to examine the necessity of the multi-head attention mechanism.

\textbullet\  $Ours@(S1_{DeepH\!E3}+S2_{GFormer})$. An implementation of our whole framework with mechanisms of DeepHE3 as well as the proposed graph Transfomer respectively serve as the two encoding and regression stages.

\begin{figure*}
	\centering	
	\includegraphics[scale=0.4]{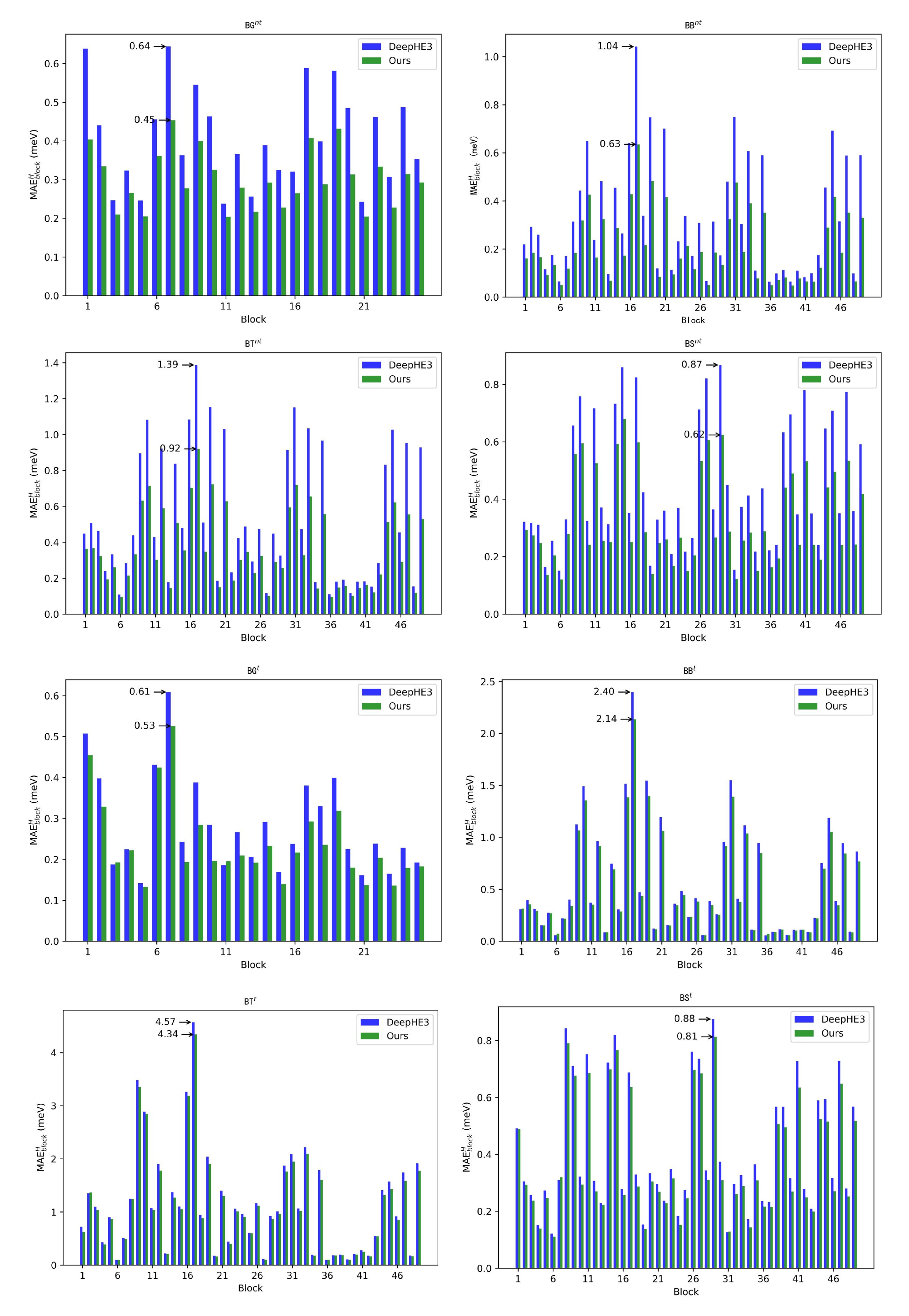}
	\caption{The MAE values (denoted as $MAE^H_{block}$) on various basic blocks of the Hamiltonian matrix in direct product state on the non-twisted (marked with superscripts $nt$) and twisted (marked with superscripts $t$) bilayer structures. The MAE values ($MAE^H_{cha\_b}$) on the Hamiltonian block where the baseline model DeepHE3 performs the worst are highlighted for comparison.}
	\label{fig:w2}
\end{figure*}

Experimental results of our complete method as well as the compared experimental settings on the six benchmark databases are presented in Tables \ref{results_mono}, \ref{results_bilayer}, and \ref{results_bilayer2}, respectively detailing the results for  monolayer structures, as well as the the results for non-twisted and twisted samples of bilayer structures. In these tables, the Mean Absolute Error (MAE) metric is used as the accuracy metric. Besides the classical MAE  metric, denoted as $MAE^H_{all}$, which measures the average error among all testing samples, we also record  $MAE^H_{cha\_s}$, the MAE for the most challenging sample where the baseline (DeepHE3) performs the worst. 	In addition to taking the Hamiltonian of each edge as a whole for accuracy statistics, since the Hamiltonian matrix in the direct product state is constituted by several basic blocks based on the angular momentum of interacting orbitals, we also conduct fine-grained accuracy statistics on these basic blocks. The MAE metrics (denoted as $MAE^{H}_{block}$) of our method and DeepHE3 on different blocks of the Hamiltonian matrix of the six structures are presented in Fig. \ref{fig:w} and  \ref{fig:w2}, where Fig. \ref{fig:w} illustrates the results on monolayer structures, while Fig. \ref{fig:w2} is dedicated to the bilayer structures. In these Figures, we specifically highlight the MAE values (denoted as $MAE^H_{cha\_b}$) on the Hamiltonian block where the baseline model DeepHE3 performs the worst for comparison on challenging blocks. All presented results are the mean values from $10$ independent repeat experiments. Since a fixed random seed is used, the  standard deviation is less than 0.008 meV and negligible.

Observed from Table \ref{results_mono}, \ref{results_bilayer}, \ref{results_bilayer2}, and Figures 1, 2, we find that the proposed hybrid approach significantly improves the prediction accuracy beyond what is achievable solely with  the priorly-equivariant mechanisms in $DeepH\!E3$ or the non-linear mechanisms in $GFormer$, demonstrating the effectiveness on leveraging the complementarity of the two categories of neural mechanisms to overcome their respective challenges comprehensively analyzed in previous sections. On one hand, the highly expressive non-linear mechanisms in the Transformer effectively compensate for the limitations in non-linear expressiveness of $DeepH\!E3$'s mechanisms, thus significantly lowering down the $MAE^H_{all}$ and $MAE^H_{cha_s}$ of $DeepHE3$, and decreasing $MAE^H_{block}$ for the vast majority of basic blocks, particularly for those blocks where $DeepHE3$ performs the worst. On the other hand, the mechanisms of $DeepH\!E3$ help the non-linear mechanisms in the Transformer to better learn equivariance from the data and reduces the difficulty for the Transformer on regressing SO(3)-equivariant targets, thus also significantly promoting the results from merely using the Transformer network. From the experimental results, it is observed that without any prior information on covariance, solely relying on the non-linear Transformer architecture  to learn SO(3)-equivariance from data is extremely challenging despite rotational augmentation, due to the complexity and high-dimensionality nature of Hamiltonians as shown in Table \ref{database-table}. And since SO(3)-equivariance is strongly linked to the intrinsic mathematical structure of Hamiltonians, the weakness on capturing SO(3)-equivariance results in inadequate modeling of Hamiltonians, leading to inaccurate predictions, especially for samples that exhibit obvious SO(3)-equivariant effects, such as twisted samples. Our framework mitigates this challenge by incorporating mechanisms with prior covariance, which helps the non-linear mechanisms in the Transformer to learn equivariance from the data and reduces the difficulty for non-linear regression of SO(3)-equivariant targets, bringing satisfactory performance of the non-linear modules. The results on twisted samples demonstrate that, our method, despite the extensive use of non-linear operators, can still capture the symmetry properties of Hamiltonians and make equivariant predictions under rotational operations. In contrast to this,  the experimental results from $S1_{DeepH\!E3}+S2_{DeepH\!E3}$ and $S1_{GFormer}+S2_{GFormer}$ shows that simply scaling up the parameters for  $DeepH\!E3$ or $GFormer$ and fine-tuning it alone only yields limited improvements. This indicates that the bottlenecks encountered by these two categories of neural mechanisms might not be fully overcome through scaling up their sizes, further highlighting the superiority and necessity of our hybrid framework.

As fine-grained ablation studies, by comparing the results of the three experimental settings, i.e., $S1_{DeepH\!E3}+S2^{-cas}_{GFormer}$, $S1_{DeepH\!E3}+S2^{-cov}_{GFormer}$, and $S1_{DeepH\!E3}+S2^{-att}_{GFormer}$, with our complete method, we could observe that the cascaded regression mechanism, the covariant feature integration mechanism, as well the multi-head attention mechanism, all contribute significantly to the performance of our method. The cascaded regression mechanism, by reducing the output space of the non-linear network, eases the difficulties on non-linear regression of Hamiltonians with SO(3)-equivariance; the covariant feature integration mechanism, through leveraging theoretical-guaranteed covariant features from DeepHE3 and geometric knowledge, successfully assists the non-linear network in learning SO(3)-equivariance; the multi-head attention mechanism, by assigning dynamic weights when fusing features, adapts to the wide variation range of geometric conditions, including both thermal deformations and twists. Under the combination of these mechanisms, networks from the two stage complement each other effectively, making our framework possess both excellent expressive capability and SO(3)-equivariant performance to achieve good results.

%\vspace{-2mm}
\section{Conclusion}
Deep learning for regressing electronic-structure Hamiltonian faces a pivotal challenge to capture SO(3)-equivariance without compromising neural expressiveness. To solve this, we propose a hybrid two-stage encoding and regression framework, where the first stage employs neural mechanisms inherent with SO(3)-equivariance properties prior to the learning process based on group theory, yielding baseline Hamiltonians with series of equivariant features assisting the subsequent stage on capturing SO(3)-equivariance. The second stage, leveraging the proposed non-linear 3D graph Transformer network for fine-grained structural analysis of 3D atomic systems, learns SO(3)-equivariant patterns from training data with the help of the first stage, while in turn, refines the initial Hamiltonian predictions via enhanced network expressiveness. Such a combination allows for accurate, generalizable Hamiltonian predictions while upholding good equivariant performance against rotational transformations. Our methodology demonstrates SOTA performance in Hamiltonian prediction, validated through six benchmark databases, showing good potentials in high-performance deep modeling of atomic systems.

\end{document}